\documentclass[12pt]{article}

\usepackage[pdftex]{graphicx}

\usepackage{epsfig,cite}

\usepackage {amsmath}

\usepackage{color}
\usepackage{rotating}
\usepackage{amssymb}

\usepackage{slashed}

\include{epsf}

\newcount\timecount

\newcount\hours \newcount\minutes  \newcount\temp \newcount\pmhours

\hours = \time

\divide\hours by 60

\temp = \hours

\multiply\temp by 60

\minutes = \time

\advance\minutes by -\temp

\def\hour{\the\hours}

\def\minute{\ifnum\minutes<10 0\the\minutes

            \else\the\minutes\fi}

\def\clock{

\ifnum\hours=0 12:\minute\ AM

\else\ifnum\hours<12 \hour:\minute\ AM

      \else\ifnum\hours=12 12:\minute\ PM

            \else\ifnum\hours>12

                 \pmhours=\hours

                 \advance\pmhours by -12

                 \the\pmhours:\minute\ PM

                 \fi

            \fi

      \fi

\fi

}

\def\monthname{\relax\ifcase\month 0/\or January\or February\or

   March\or April\or May\or June\or July\or August\or September\or

   October\or November\or December\else\number\month/\fi}

\def\bold#1{\setbox0=\hbox{$#1$}%

     \kern-.025em\copy0\kern-\wd0

     \kern.05em\copy0\kern-\wd0

     \kern-.025em\raise.0433em\box0 }

\def\beq{\begin{equation}}

\def\eeq{\end{equation}}

\def\ga{\mathrel{\raise.3ex\hbox{$>$\kern-.75em\lower1ex\hbox{$\sim$}}}}

\def\la{\mathrel{\raise.3ex\hbox{$<$\kern-.75em\lower1ex\hbox{$\sim$}}}}

\def\gev{{\rm \, Ge\kern-0.125em V}}

\def\tev{{\rm \, Te\kern-0.125em V}}

\def\gyr{{\rm \, G\kern-0.125em yr}}

\def\gappeq{\mathrel{\rlap {\raise.5ex\hbox{$>$}}

{\lower.5ex\hbox{$\sim$}}}}

\def\lappeq{\mathrel{\rlap{\raise.5ex\hbox{$<$}}

{\lower.5ex\hbox{$\sim$}}}}

\def\Toprel#1\over#2{\mathrel{\mathop{#2}\limits^{#1}}}

\def\m12{m_{1\!/2}}

\def\bea{\begin{eqnarray}}

\def\eea{\end{eqnarray}}

\def\beqar{\begin{eqnarray}}

\def\eeqar{\end{eqnarray}}

\def\m{{\cal m}}

\begin{document}

\begin{titlepage}

\pagestyle{empty}

\baselineskip=21pt

%\rightline{\tt astro-ph/yymmnnn}

\rightline{KCL-PH-TH/2012-16, LCTS/2012-09, CERN-PH-TH/2012-080}

\vskip 1in

\begin{center}

{\large {\bf Global Analysis of Experimental Constraints on \\ a Possible Higgs-Like Particle with Mass $\sim 125$~GeV}}

\end{center}

\begin{center}

\vskip 0.2in

 {\bf John~Ellis}$^{1,2}$
and {\bf Tevong~You}$^3$

\vskip 0.1in

{\small {\it

$^1${Theoretical Particle Physics and Cosmology Group, Physics Department, \\
King's College London, London WC2R 2LS, UK}\\

$^2${TH Division, Physics Department, CERN, CH-1211 Geneva 23, Switzerland}\\

$^3${High Energy Physics Group, Blackett Laboratory, Imperial College, Prince Consort Road, London SW7 2AZ, UK}\\
}}

\vskip 0.2in

{\bf Abstract}

\end{center}

\baselineskip=18pt \noindent

%%%%%%%%%%%%%%%%%%%%%%%%%%%%%%%%%%%%%%%%%%%%%%%%%

{\small
We perform a global analysis of the constraints on a possible Higgs-like particle
with mass $\sim 125$~GeV that are provided by the ATLAS, CDF, CMS and D0
experiments. We combine the available constraints on possible deviations from
the Standard Model Higgs couplings to massive vector bosons and to fermions,
considering also the possibilities of non-standard loop-induced couplings to
photon and gluon pairs. We analyze the combined constraints on pseudo-dilaton
scenarios and on some other scenarios in which the possible new particle is
identified as a pseudo-Nambu-Goldstone boson in a composite electroweak
symmetry-breaking sector.}

%%%%%%%%%%%%%%%%%%%%%%%%%%%%%%%%%%%%%%%%%%%%%%%%

\vfill

\leftline{%CERN-PH-TH/2011-xxx, KCL-PH-TH/2011-xxx, LCTS-2011-yy, 
March 2012}

\end{titlepage}

\baselineskip=18pt

%%%%%%%%%%%%%%%%%%%%%%%%%%%%%%%%%%%%%%%%%%%%%%%%%%

\section{Introduction}

The LHC experiments ATLAS and CMS have reported evidence for a possible
new particle with mass $\sim 125$~GeV~\cite{LHCcombined,ZZsearch, bbbarsearch, CMSdiphotonsearch, ATLASdiphotonsearch, ATLASWWsearch, CMSWWsearch, CMSATLAStautausearch, CMStautaumumu, CMSWHsearch,CMSprevdiphotonsearch}, 
whose existence is also consistent with
data from the Tevatron collider experiments CDF and D0 \cite{Tevatronsearch}. It seems quite likely that the new particle would have spin
zero, since it apparently couples to $\gamma \gamma$ and hence cannot have
spin one, and the selections of candidate $W W^* \to \ell^+ \ell^-\nu {\bar \nu}$ final states assume
that the charged lepton momenta are
correlated as in the decays of a spin-zero particle such as a Higgs boson~\cite{Higgsboson}, 
whereas a spin-two particle would yield quite different
correlations \cite{JEDSHspin2}. This state is therefore a very plausible Higgs candidate, but could
be a harbinger of a more complicated, possible composite, electroweak symmetry-breaking sector.

Inventive theorists have proposed many such alternative
scenarios with signatures somewhat different from those of the Standard Model
Higgs boson, and it will be important to optimize the use of the sparse initial data to best
distinguish between them. Many previous papers have proposed phenomenological
frameworks that generalize the couplings of the Standard Model Higgs boson \cite{generalcouplings, ac}, and several
papers have already used such frameworks to analyze the possible couplings of the
$\sim 125$~GeV Higgs candidate \cite{Contino,mh125papers}, most recently using the data released by ATLAS, CDF,
CMS and D0 for the March 2012 Moriond conference~\cite{Moriond}.

In this paper we set up a calculational tool for analyzing and combining the constraints
provided by the various experimental measurements, treating as independent parameters
the strengths of the couplings to massive vector bosons and to fermions, and allowing for
the possibilities of non-standard loop-induced couplings to photon and gluon pairs.
Within this general framework, we analyze the experimental constraints on some specific
alternatives to the Standard Model Higgs boson. 

These include the pseudo-dilaton, the pseudo-Nambu-Goldstone boson of a near-conformal 
strongly-interacting sector with vacuum condensates that break both scale and
electroweak symmetry \cite{pseudoDG, Yamawaki}. In this model, the tree-level couplings to all massive
Standard Model particles, bosons and fermions, are rescaled by the same universal
factor relative those of the Higgs boson in the Standard Model. In addition, there may
be extra contributions to the loop couplings to $\gamma \gamma$ and $gg$,
one possibility being that QCD and QED also become almost conformal~\cite{pseudoDG}. Neglecting
this possibility for QCD, we find that the combined world data on the $\sim 125$~GeV state favour
a universal rescaling factor that is close to unity. On the other hand, if QCD is near-conformal,
favoured values of the rescaling factor are substantially less than unity.

Other possibilities are provided by models in which the `Higgs' is a composite pseudo-Nambu-Goldstone
boson of some higher chiral symmetry that is broken down to the SU(2) $\times$ SU(2) $\to$
SU(2) of the Standard Model Higgs sector \cite{compositeHiggs,minimalCompositeHiggs,Espinosaetal,RattazziZurich}~\footnote{Similar phenomenology occurs
in some radion models~\cite{radion}.}. In some of these models the `Higgs' couplings
to massive vector bosons and to fermions are rescaled differently, with extreme cases
being fermiophobic~\cite{fermiophobic} and gaugephobic~\cite{gaugephobic} models. We analyze these possibilities,
comparing the qualities of their fits to that of the Standard Model Higgs boson and the 
pseudo-dilaton. We find that improved fits are possible in alternative models
in which the relative signs of the `Higgs'-fermion and -boson couplings are opposite from the Standard Model,
but the improvement is not sufficient to warrant discarding the Standard Model Higgs or the
dilaton, whereas fermiophobic and (particularly) gaugephobic models are disfavoured. 

The layout of this paper is as follows. In Section~2 we review the phenomenological
framework we employ, in Section~3 we describe our calculational procedure, and in
Section~4 we describe the data set we use. In Section~5 we present our results, treating as particular cases
the pseudo-dilaton scenario and other pseudo-Nambu-Goldstone
boson scenarios, as well as fermiophobic and gaugephobic models. In each
single-parameter scenario, we present the local $p$-value (likelihood) as a function of
its relevant parameter  Finally, in Section~6 we summarize our conclusions and discuss
prospects for clarifying the nature of the possible state with mass $\sim 125$~GeV.

\section{Phenomenological Framework}

We work within the framework of the following nonlinear low-energy effective Lagrangian for the
electroweak symmetry-breaking sector~\cite{ac,pseudoDG}, see also~\cite{pionL,dilatonL}:
\bea
{\cal L}_{eff} \; & = & \; \frac{v^2}{4} {\rm Tr} \left(D_\mu U D^\mu U^\dagger \right) \times \left[ 1 + 2 a \frac{h}{v} + b \frac{h^2}{v^2} + \dots \right] \nonumber \\
& - & \frac{v}{\sqrt{2}} \left( {\bar u}_L, {\bar d}_L \right) U \left[ 1 + c \frac{h}{v} + \dots \right]
 \left( \begin{matrix} \lambda_u u_R \\  \lambda_d d_R \end{matrix} \right) + h.c.
\label{effL}
\eea
where $U$ is a unitary $2 \times 2$ matrix parametrizing the three Nambu-Goldstone fields
that are `eaten' by the $W^\pm$ and $Z^0$, giving them masses, $v \sim 246$~GeV is the
conventional electroweak symmetry-breaking scale, $h$ is a field describing the possible state with mass
$\sim 125$~GeV, and $a$ and $c$ parametrize the deviations of its couplings to massive
vector bosons and to fermions, respectively, from those of the Higgs boson in the Standard Model.
In addition to the terms in (\ref{effL}), the phenomenology of a Higgs-like state $h$ also
depends on the loop-induced dimension-5 couplings to $gg$ and $\gamma \gamma$~\cite{EGN}:
\begin{equation}
{\cal L}_{\Delta} \; = \; - \left[ \frac{\alpha_s}{8 \pi} b_s 
G_{a \mu \nu} G_a^{\mu \nu} + \frac{\alpha_{em}}{8 \pi} b_{em} F_{\mu \nu} F^{\mu \nu} \right] \left(\frac{h}{V}\right) .
\label{triangles}
\end{equation}
In the Standard Model, only the top quark makes a significant contribution to the coefficient $b_s$, whereas both the top
quark and $W^\pm$ contribute to $b_{em}$. In extensions of the Standard Model,
other heavy particles may contribute to both coefficients, as we discuss below.

We will be particularly interested in the case where $h$ is the pseudo-Goldstone boson of
approximate scale symmetry, i.e., a pseudo-dilaton of a near-conformal electroweak symmetry-breaking
sector. In this case, the square-bracketed factors in (\ref{effL}) may be written in the forms \cite{dilatonL}:
\beq
\left[ \dots \right]_1 \; = \; \left[ \left( \frac{\chi}{V} \right)^2 \right] \; , \;
\left[ \dots \right]_2 \; = \; \left[ \frac{\chi}{V} \right] 
\label{dilatonfactors}
\eeq
where $\chi$ is the dilaton field, assumed to have a v.e.v. $V$, and we may write $\chi \equiv V + h$.
In this pseudo-dilaton scenario, we have
\beq
a \; = \; c \; = \frac{v}{V} .
\label{dilatonratio}
\eeq
In addition to this most economical pseudo-Goldstone boson scenario, we also consider
scenarios in which $h$ is interpreted as a pseudo-Goldstone boson appearing when some
higher-order chiral symmetry is broken down to the SU(2) $\times$ SU(2) of the Standard Model
Higgs sector. One class of such composite models has an SO(5)/SO(4) structure \cite{minimalCompositeHiggs}, within which
the Standard Model fermions may be embedded in spinorial representations of SO(5), the
MCHM4 model, or in fundamental representations, the MCHM5 model \cite{ac,Espinosaetal,RattazziZurich,mh125papers}. In the MCHM4 model
one has
\beq
a \; = \; c \; = \; \sqrt{1 - \xi} ,
\label{MCHM4}
\eeq
where $\xi \equiv (v/f)^2$ with $f$ a compositeness scale. Clearly, constraints on the possible values of
$a = c$ in the pseudo-dilaton scenario may be rephrased as constraints on $\xi$ in the MCHM4 model.
On the other hand, in the MCHM5 model one has
\beq
a \; = \; \sqrt{1 - \xi} \; , \; c \; = \; \frac{1 - 2 \xi}{\sqrt{1 - \xi}} .
\label{MCHM5}
\eeq
This reduces to the Standard Model as $\xi \to 0$, to a specific fermiophobic scenario with $a = \sqrt{3}/2$
in the limit $\xi \to 1/2$, to an anti-dilaton model with $a = - c = 1/\sqrt{3}$ when $\xi = 2/3$, and to a
gaugephobic model when $\xi \to 1$.

\section{Calculational Procedure}

The deviations of the $h$ couplings from those of the Standard Model Higgs boson, 
parametrized by $a$ and $c$, factorize out of the Standard Model production cross-sections and decay widths,
yielding the following rescaling factors $R\equiv\sigma/\sigma_{\text{SM}}$ for gluon-gluon fusion, vector boson fusion, associated production and Higgsstrahlung production mechanisms respectively: 
\begin{equation} 
	R_{gg} = c^2	\quad , \quad R_{\text{VBF}} = a^2 \quad , \quad R_{\text{ap}} = a^2 \quad , \quad R_{\text{hs}} = c^2  \quad . 
\end{equation}
Assuming that gluon-gluon fusion and vector boson fusion (VBF) dominate over the other processes, 
one may combine their respective rescaling factors and cut efficiencies $\xi_{\text{gg,VBF}}$ to obtain 
a total production rescaling factor 
\begin{equation}
	R_{\text{prod}} = \frac{\xi_{gg}F_{gg}R_{gg} + \xi_{\text{VBF}}(1-F_{gg})R_{\text{VBF}}}{\xi_{gg}F_{gg} + \xi_{\text{VBF}}(1-F_{gg})}		\quad ,
\end{equation}
where $F_{gg} \equiv \sigma^{\text{SM}}_{gg}/\sigma^\text{SM}_\text{tot}$. 

Similarly the rescaling of the decay widths $R\equiv\Gamma/\Gamma_{\text{SM}}$ to massive vector bosons, fermions and photons are given, respectively, by
\begin{equation}
	R_{VV} = a^2	\quad , \quad	R_{\bar{f}f} = c^2	\quad , \quad	R_{\gamma\gamma} = \frac{(-\frac{8}{3}cF_t + aF_w)^2}{(-\frac{8}{3}F_t+F_w)^2}	\quad , 
\label{loopfactors}
\end{equation}
where 
\begin{align*}
	F_t &= \tau_t\left[1+(1-\tau_t)f(\tau_t)\right]	\quad ,	\\ 
	F_w &= 2 + 3\tau_w\left[1+(2-\tau_w)f(\tau_w) \right]		\quad , 	\\
	f(\tau) &= \left\{ \begin{array}{c c} 
		\left(\arcsin\sqrt{\frac{1}{\tau}}\right)^2 	\quad\quad &\tau \geq 1 \\ 
		-\frac{1}{4}\left(\log\frac{1+\sqrt{1-\tau}}{1-\sqrt{1-\tau}} - i\pi\right)^2	\quad\quad &\tau < 1
	\end{array} 	\right.	\quad ,
\end{align*}
and $\tau_{t,w} \equiv  4m_{t,w}^2/m_h^2$. The factors entering in $R_{\gamma\gamma}$ 
arise from the top-quark and $W$-boson triangle loops. The principal dependences of the
different Higgs-like signals on the rescaling factors $(a, c)$ are summarized in Table~\ref{table:channeleffparamssensitivity}.

\begin{table}[h!]
	\center
	\begin{tabular}{ | c | c | c | c | c | c | c |}
		\hline
		 & \multicolumn{3}{|c|}{Production sensitive to} & \multicolumn{3}{|c|}{Decay sensitive to} \\ 
		channel & \quad $a$ \quad & \quad $c$ \quad & $b_{s}$ & \quad $a$ \quad & \quad $c$ \quad & $b_{em}$ \\ \hline
		$\gamma\gamma$ & $\checkmark$ & $\checkmark$ & $\checkmark$ & $\checkmark$ & $\checkmark$ & $\checkmark$ \\ \hline
		$\gamma\gamma$ VBF & $\checkmark$ & $\times$ &  $\times$ & $\checkmark$ & $\checkmark$ & $\checkmark$ \\ \hline
		WW & $\checkmark$ & $\checkmark$ & $\checkmark$ & $\checkmark$ & $\times$ & $\times$ \\ \hline 
		WW 2-jet & $\checkmark$ & $\times$ & $\times$ & $\checkmark$ & $\times$ & $\times$ \\ \hline
		WW 0,1-jet & $\times$ & $\checkmark$ & $\checkmark$ & $\checkmark$ & $\times$  &  $\times$ \\ \hline
		$b\bar{b}$ & $\checkmark$ & $\times$ & $\times$ & $\times$ & $\checkmark$ & $\times$ \\ \hline
		ZZ & $\checkmark$ & $\checkmark$& $\checkmark$ & $\checkmark$ & $\times$ & $\times$ \\ \hline
		$\tau\tau$ & $\checkmark$ & $\checkmark$ & $\checkmark$ & $\times$ & $\checkmark$ & $\times$ \\ \hline 
		$\tau\tau\to\mu\mu$ & $\checkmark$ & $\checkmark$ & $\checkmark$ & $\times$ & $\checkmark$ & $\times$ \\ \hline 
	\end{tabular}
	\caption{\it Dominant dependences on the model parameters for the channels and sub-channels discussed
	in this paper, where $a$ and $c$ control the strength of the scalar coupling to the massive gauge bosons and
	to fermions, respectively, and $b_s$, $b_\text{em}$ are the coefficients of the dimension-5 term 
	coupling the scalar to the massless gauge bosons. The latter coefficients are important factors in pseudo-dilaton phenomenology.  }
	\label{table:channeleffparamssensitivity}
\end{table}

In some scenarios there may be additional loop contributions
due to new heavy particles that should also be taken into account, an example being the pseudo-dilaton scenario. 
If QCD and QED are embedded in the conformal sector, the gluon-gluon fusion and diphoton decay rates are rescaled by factors
related to the trace anomaly \cite{traceanomaly}:
\begin{equation}
	R_{gg} = \frac{(-\frac{v}{V}b_s + cF_t)^2}{F_t^2} 	\quad , \quad	R_{\gamma\gamma} = \frac{(-\frac{v}{V}b_{em} -\frac{8}{3}cF_t + aF_w)^2}{(-\frac{8}{3}F_t+F_w)^2}	\quad ,
\end{equation}
where 
\begin{align*}
	b_{s} &= \left\{ \begin{array}{c c} 
		11- \frac{10}{3} \quad & m_h < m_t	\\
		11- 4 \quad & m_h \geq m_t  
	\end{array}	\right.	\quad  \\
	b_{em} &= \left\{ \begin{array}{c c} 
		-\frac{17}{9} \quad & m_W <  m_h < m_t		\\
		-\frac{11}{3} \quad & m_h > m_t	
	\end{array}	\right. 	\quad
	 ,
\end{align*}
and the forms of the loop-induced dimension-5 terms in the effective Lagrangian and the definitions of the coefficients
$b_s$ and $b_{em}$ were given in (\ref{triangles}).

The signal strength modification factor $\mu^i \equiv n^i_s / (n^i_s)^\text{SM}$ in any given channel $i$ may be calculated
by combining the production and decay rescalings: $R \equiv R^i_{\text{prod}}\cdot (R^i_\text{decay}/R_\text{tot.})$. 
Experimental collaborations typically report the expected and observed 95\% CL limits on $\mu$ from experimental 
searches for the Standard Model Higgs boson. In the absence of more detailed experimental information on the
likelihood function in each channel, we adopt the following approximate procedure to re-interpret these results for 
different signal strength modifiers and hence in the $a$ and $c$ parameter space~\cite{Contino}. 

The underlying likelihood $p(n_\text{obs} | \mu n_s^\text{SM} + n_b)$ is assumed to obey a Poisson distribution.
In the Gaussian limit of a large number of observed events, assuming small fluctuations with respect to the 
background and negligible systematic errors~\footnote{See Ref. \cite{Contino} for a more precise definition of these assumptions.}, 
one can use the approximation $\sigma_{\text{obs}} \simeq \sigma_{\text{exp}} = \mu^{95\%}_\text{exp}/1.96$  for the standard deviation
to solve for the central value $\bar{\mu}$ in the equation~\footnote{This is most easily done numerically using the Gauss error function.}:
\begin{equation}
	\frac{\int^{\mu^{95\%_\text{obs}}}_0 e^{-\frac{(\mu - \bar{\mu})^2}{2\sigma^2_\text{obs}}} d\mu}{\int^\infty_0 e^{-\frac{(\mu - \bar{\mu})^2}{2\sigma^2_\text{obs}}} d\mu} = 0.95	\quad .
\end{equation} 
%
%The value of $1.96$ in the denominator of (\ref{eq:sigmaapprox}) was obtained by a similar relation for the expected limit. 
The posterior probability density function is then given by 
\begin{equation}
	p(\mu | n_\text{obs}) = p(n_\text{obs} | \mu n_s^\text{SM} + n_b)\cdot\pi(\mu) \approx \frac{1}{\sqrt{2\pi\sigma_\text{obs}^2}}e^{-\frac{(\mu-\bar{\mu})^2}{2\sigma_\text{obs}^2}}	\quad ,
\end{equation}
with $\pi(\mu)$ generally assumed {\it a priori} to be flat within the range of interest and zero outside. As discussed below,
we find that this method yields results within 20\% of official combinations, which is sufficient for our purpose.

\section{Experimental Data Set} 

To reconstruct the likelihood we used the latest available information based on up to $\sim 5$/fb of LHC data
and $\sim 10$/fb of Tevatron data per experiment, as follows.

\begin{enumerate}
	\item The CMS and ATLAS searches in the channel $h \to ZZ \to 4 \ell^\pm$ are treated as inclusive~\cite{ZZsearch}.
	\item The CMS, ATLAS and Tevatron searches in the $h \to \bar{b}b$ channel are assumed to be dominated by associated production~\cite{bbbarsearch, Tevatronsearch}.
	\item The CMS search in the $h \to \gamma \gamma$ channel is split into five sub-channels, 
	whose likelihoods we estimate from the quoted 
	central values and one-sigma error bars given  in~\cite{CMSdiphotonsearch}. 
	The dijet-tagged sub-channel is dominated by VBF, with the efficiencies for $m_h=120$ GeV quoted as 
	$\xi_{gg} = 0.05$, $\xi_\text{VBF}=0.15$ in~\cite{CMSprevdiphotonsearch}, whereas the the other sub-channels are 
	dominated by $gg$ fusion and combined. The ATLAS diphoton search is treated as inclusive~\cite{ATLASdiphotonsearch}.
	\item The ATLAS and Tevatron searches in the $h \to W^+W^-$ channel are treated as inclusive~\cite{ATLASWWsearch, Tevatronsearch}.
	\item The CMS $h \to W^+W^-$ search results include detailed tables of the numbers of events in~\cite{CMSWWsearch}
	in 0-jet, 1-jet and dijet sub-channels for various choices of $m_h$. We expect that the numbers for $m_h \sim 125$~GeV
	would lie between those for 120 and 130~GeV. Conservatively, we use the numbers given for $m_h=120$ GeV 
	to reconstruct the likelihood, as they yield a weaker constraint than the numbers for 130~GeV.
	We further assume that the dijet sub-channel is dominated by VBF.
	\item The CMS and ATLAS $\tau\tau$ searches are treated as inclusive \cite{CMSATLAStautausearch}. For CMS we also include an additional search in the $h\to\tau\tau\to\mu\mu$ channel \cite{CMStautaumumu}.
\end{enumerate}

We omit two searches using associated $WH$ production \cite{CMSWHsearch}, 
which currently have limited statistics and sensitivity.
The cut efficiencies are assumed to be independent of $(a, c)$ for all channels other than the 
diphoton dijet-tagged sub-channel. We do not make any 
allowances for non-Gaussian systematic errors or correlations in our study. 
Where possible, we have checked that our combinations agree with the official combinations to within $\sim 20\%$, 
e.g., for the combination of the CMS diphoton sub-channels and for the combination of CMS channels.
However, it should be noted that this approach is inherently limited by the available information, 
and care should be taken not to over-interpret the results. For this reason, 
we would welcome the release by the experimental collaborations of more information 
about the likelihood functions for each sub-channel analyzed \cite{HouchesRecommendations}.

\section{Results}

\begin{figure}
\vskip 0.5in
\vspace*{-1.1in}
%\hspace*{-.70in}
\begin{minipage}{8in}
\includegraphics[height=6cm]{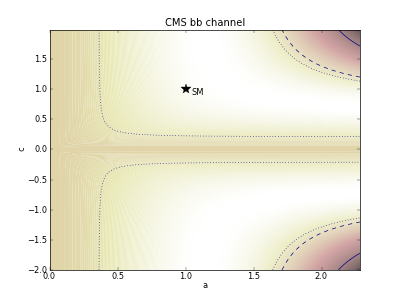}
\hspace*{0.2in}
\includegraphics[height=6cm]{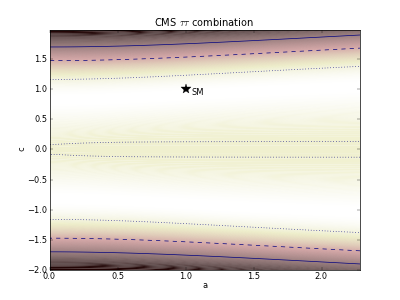}
\hfill
\end{minipage}
\begin{minipage}{8in}
\includegraphics[height=6cm]{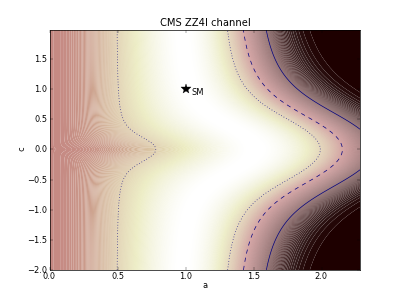}
\hspace*{0.2in}
\includegraphics[height=6cm]{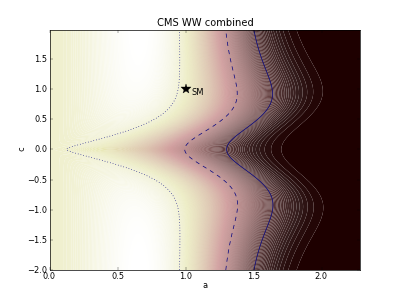}
\hfill
\end{minipage}
\begin{minipage}{8in}
\includegraphics[height=6cm]{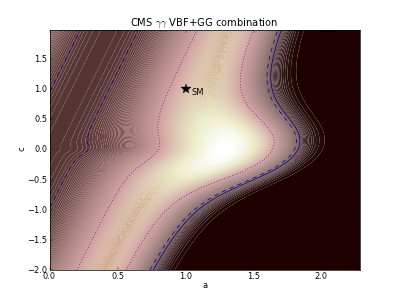}
\hspace*{0.2in}
\includegraphics[height=6cm]{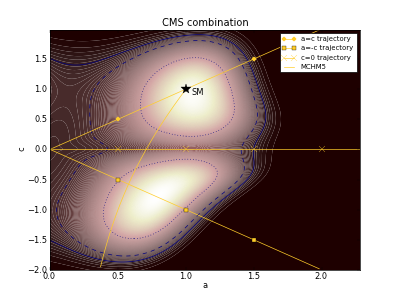}
\hfill
\end{minipage}
\caption{
{\it
CMS constraints on the couplings $(a, c)$ of a possible Higgs-like particle $h$ with mass $\sim 125$~GeV
arising from the (top left) ${\bar b} b$, (top right) $\tau^+ \tau^-$,
(centre left)  $Z Z^*$, (centre right) $W W^*$ and (bottom left) $\gamma \gamma$ final states. 
The bottom right panel displays the combination of these CMS constraints, together with lines
representing the pseudo-dilaton, anti-dilaton, fermiophobic and MCHM5 scenarios.
In these and subsequent analogous plots
the most likely regions have the lightest shading,
the dotted lines are 68\% CL contours, the dashed lines are 95\% CL contours, and the solid lines
are 99\% CL contours.}} 
\label{fig:CMS} 
\end{figure}

We first discuss the constraints on $a, c$ that are imposed by the CMS data released in March 2012.
The panels in Fig.~\ref{fig:CMS} display the constraints from the (top left) ${\bar b} b$, (top right) $\tau^+ \tau^-$,
(centre left)  $Z Z^*$, (centre right) $W W^*$ and (bottom left) $\gamma \gamma$ final states. Also shown in
the bottom right panel is the combination of these CMS constraints. In these and subsequent analogous plots
the most likely regions have the lightest shading,
the dotted lines are 68\% CL contours, the dashed lines are 95\% CL contours, and the solid lines
are 99\% CL contours. 

We see in the top panels of Fig.~\ref{fig:CMS} that the
${\bar b} b$ channel depends on both $a$ and $c$ and that the $\tau^+ \tau^-$ measurements are mainly sensitive to $c$, as one would expect,
with a very weak dependence on $a$ induced by the subdominant vector-boson-fusion (VBF)
production mechanism. On the other hand, we see in the centre left panel 
that the $Z Z^*$ channel is quite sensitive to both $a$ and $c$,
but is insensitive to the sign of $c$. 

\begin{figure}
\vskip 0.5in
\vspace*{-0.75in}
%\hspace*{-.70in}
\begin{minipage}{8in}
\includegraphics[height=6cm]{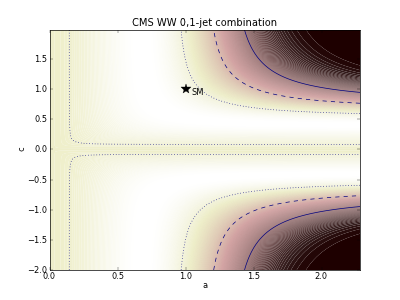}
\hspace*{0.2in}
\includegraphics[height=6cm]{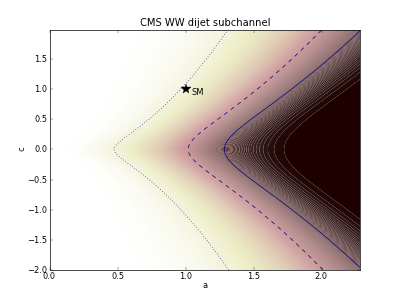}
\hfill
\end{minipage}
\caption{
{\it
Breakdown of the CMS constraints on the couplings $(a, c)$ of a possible Higgs-like $h \to WW^*$ signal with mass $\sim 125$~GeV
arising from production via (left) $gg$ fusion and (right) vector-boson-fusion (VBF).}} 
\label{fig:CMSWW} 
\end{figure}

The $W W^*$ channel in the centre right panel is also sensitive to both $a$ and $|c|$,
but in a completely different way. The decay branching ratios for $WW^*$ and $ZZ^*$ are closely
related, and the favouring of small $a$ by the $WW^*$ constraint reflects the (not very significant)
suppression of the CMS $WW^*$ signal relative to the $ZZ^*$ signal. However, in the case of the $WW^*$
channel, CMS provides additional information via separate analyses of the $WW^*$ + 0, 1, 2-jet
channels, which provide some discrimination between the $gg$ fusion and VBF production mechanisms,
as shown in the left and right panels of Fig.~\ref{fig:CMSWW}, respectively~\footnote{CMS provide this
breakdown for the cases $m_h = 120$ and 130~GeV~\cite{CMSWWsearch}, 
rather than for the favoured signal region at $m_h \sim 125$~GeV.
Conservatively, we apply here the $m_h = 120$~GeV version of the $WW^*$ constraint, which is weaker than at
larger masses.}. We see, in particular, that the VBF constraint (right panel) disfavours the fermiophobic
limit $c \to 0$. 

\begin{figure}
\vskip 0.5in
\vspace*{-0.75in}
%\hspace*{-.70in}
\begin{minipage}{8in}
\includegraphics[height=6cm]{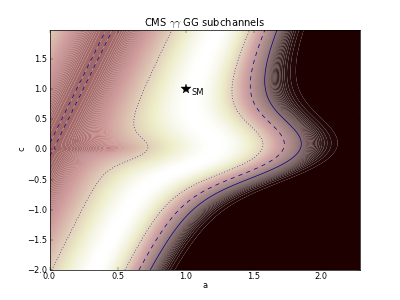}
\hspace*{0.2in}
\includegraphics[height=6cm]{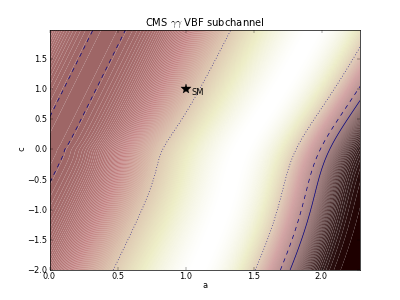}
\hfill
\end{minipage}
\caption{
{\it
Breakdown of the CMS constraints on the couplings $(a, c)$ of a possible Higgs-like $h \to \gamma \gamma$ 
signal with mass $\sim 125$~GeV arising from production via (left) $gg$ fusion and (right) vector-boson-fusion (VBF).}} 
\label{fig:CMSgammagamma} 
\end{figure}

Fig.~\ref{fig:CMSgammagamma}
displays in its left and right panels an analogous breakdown of the $gg$ and VBF constraints on the
$h \to \gamma \gamma$ signal. We see that in this case the large-$a$ possibility is actually
favoured, since the VBF-enhanced $h \to \gamma \gamma$ signal is relatively strong.
The $gg$ and VBF constraints on $h \to \gamma \gamma$ are combined in the bottom left panel of Fig.~\ref{fig:CMS},
just as the $gg$ and VBF constraints on the $h \to WW^*$ signal are combined in the centre right
panel of Fig.~\ref{fig:CMS}. We note that the $h \to \gamma \gamma$ constraint is not symmetric
between positive and negative $c$, because the loop-induced $h \to \gamma \gamma$ decay amplitude
is sensitive to the relative sign of $a$ and $c$ through the interference between ${\bar t} t$ and $W^+ W^-$
loops. We also note that the best fit to the CMS $\gamma \gamma$ data is in the fermiophobic region.

\begin{figure}
\vskip 0.5in
\vspace*{-0.75in}
%\hspace*{-.70in}
\begin{minipage}{8in}
\includegraphics[height=6cm]{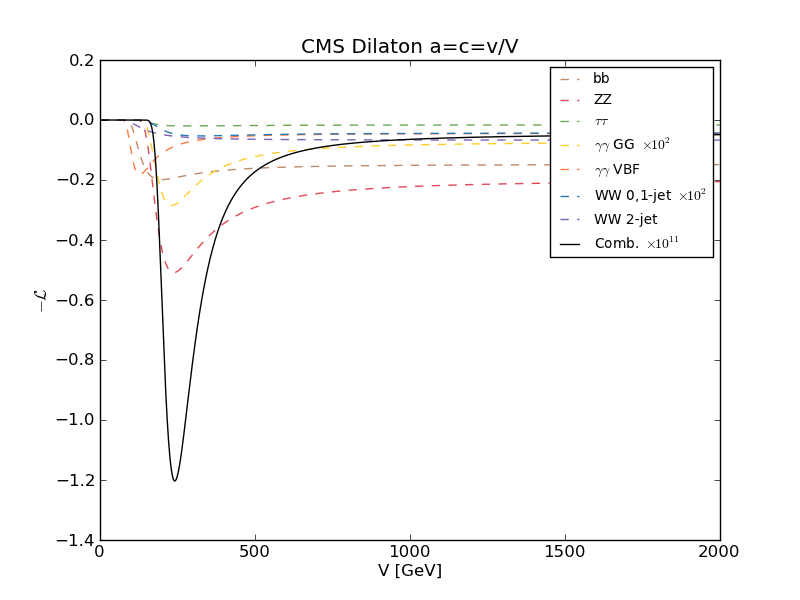}
\hspace*{-0.17in}
\includegraphics[height=6cm]{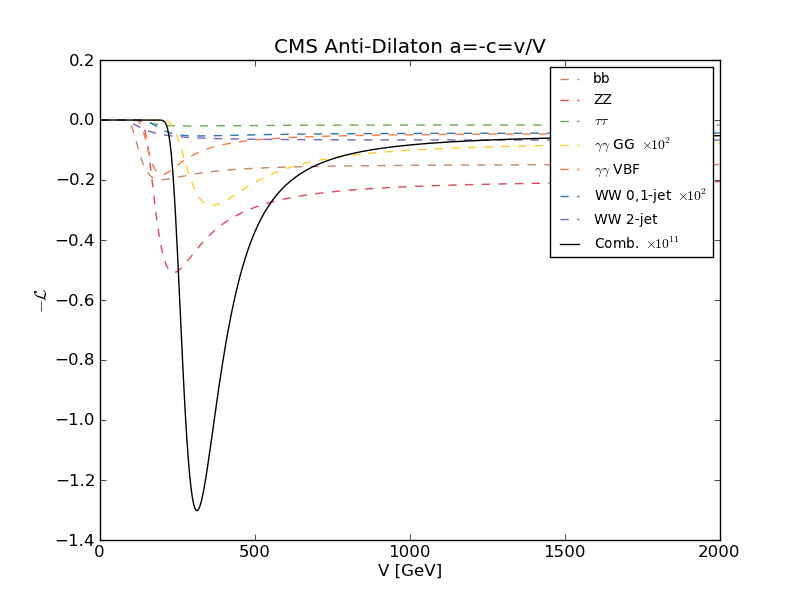}
\hfill
\end{minipage}
\begin{minipage}{8in}
\includegraphics[height=6cm]{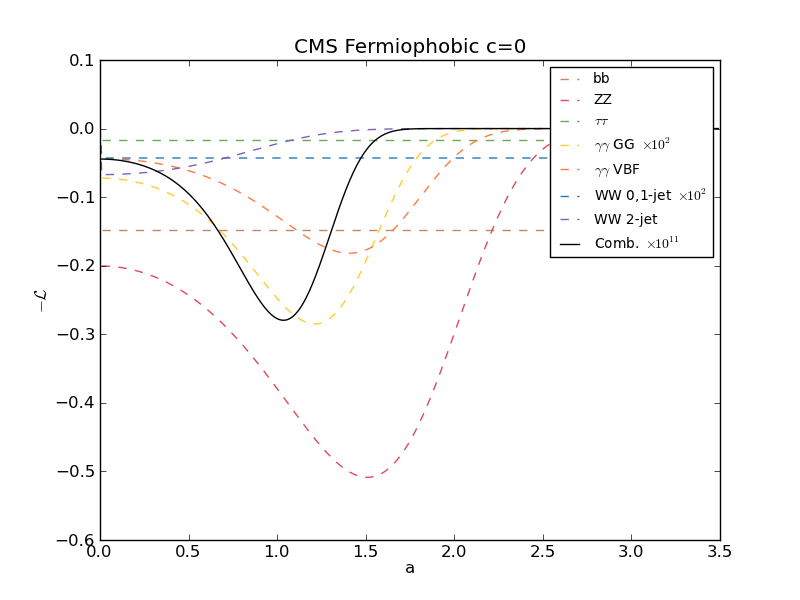}
\hspace*{-0.17in}
\includegraphics[height=6cm]{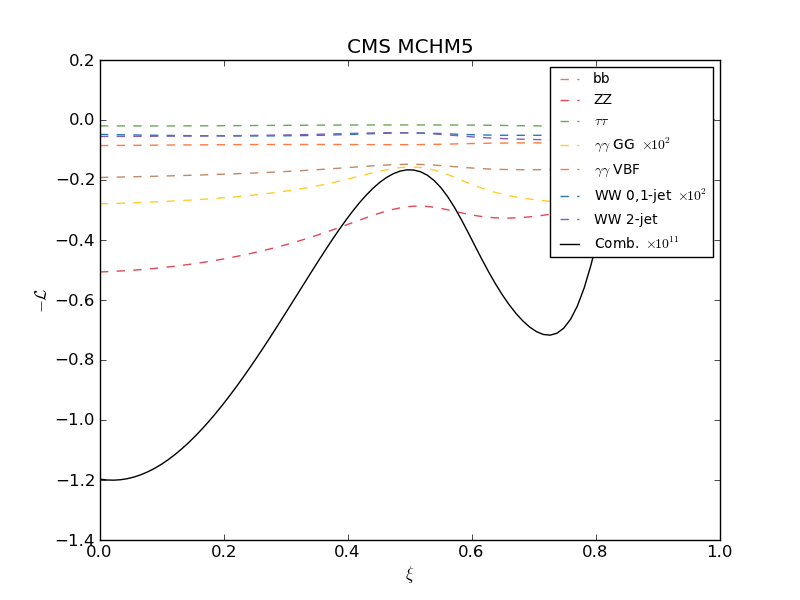}
\hfill
\end{minipage}
\caption{
{\it
The $p$-values of global fits to the CMS constraints along (upper left) the pseudo-dilaton line $a = c$,
(upper right) the `anti-dilaton' line $a = -c$, (lower left) the fermiophobic line $c = 0$, and
(lower right) the MCHM5 line $a c = 1 - 2\sqrt{1 - a^2}$.
}}
\label{fig:CMSpvalues} 
\end{figure}

Turning now to the bottom right panel of Fig.~\ref{fig:CMS} that displays the combination of the
CMS constraints on the $h$ couplings, we see that the combination is almost symmetric between
$c > 0$ and $c < 0$, with the latter being slightly favoured, and that the fermiophobic
case $c = 0$ is quite strongly disfavoured. The diagonal lines in this combination panel
represent the pseudo-dilaton case with $a = c$ and a hypothetical `anti-dilaton' model with $a = - c$.
The curved line $a c = 1 - 2\sqrt{1 - a^2}$ represents the MCHM5 model (\ref{MCHM5}) that is
parametrized by $\xi$. In Fig.~\ref{fig:CMSpvalues} we show the local $p$-values found for
the CMS data along (upper left) the dilaton 
and (upper right) antidilaton lines, (lower left) the MCHM5 line, and (lower right) the line $c = 0$ corresponding to fermiophobic models.
We see the best fit to the pseudo-dilaton model has $a = c = v/V$ close to unity, corresponding to the
$h$ couplings being very simllar to those of a Standard Model Higgs boson. The local $p$-value for
the `anti-dilaton' model is maximized for $a = - c$ slightly less than unity, and is slightly favoured over
the dilaton scenario, though not to a significant extent. On the other hand, the fermiophobic scenario
is significantly disfavoured. In the MCHM5 case, we see a global Standard-Model-like minimum at $\xi = 0$
and a local minimum in the anti-dilaton region (with sub-optimal coupling strengths).

\begin{figure}
\vskip 0.5in
\vspace*{-1.1in}
%\hspace*{-.70in}
\begin{minipage}{8in}
\includegraphics[height=6cm]{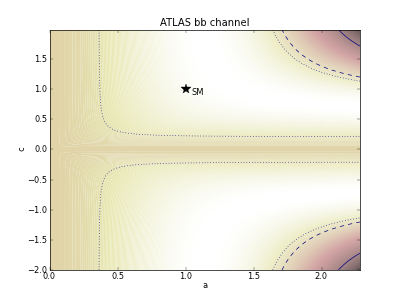}
\hspace*{0.2in}
\includegraphics[height=6cm]{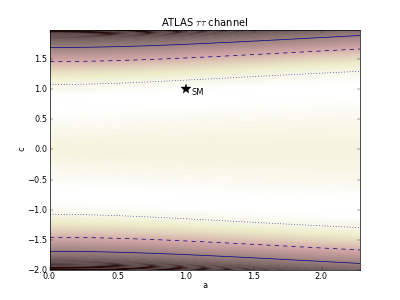}
\hfill
\end{minipage}
\begin{minipage}{8in}
\includegraphics[height=6cm]{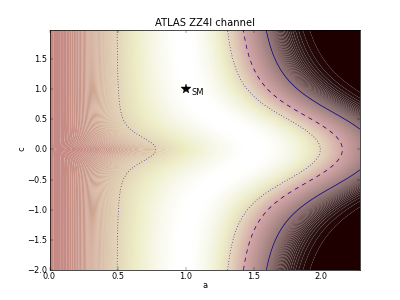}
\hspace*{0.2in}
\includegraphics[height=6cm]{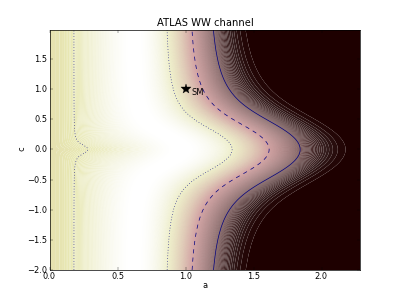}
\hfill
\end{minipage}
\begin{minipage}{8in}
\includegraphics[height=6cm]{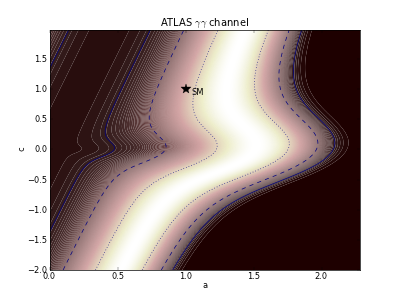}
\hspace*{0.2in}
\includegraphics[height=6cm]{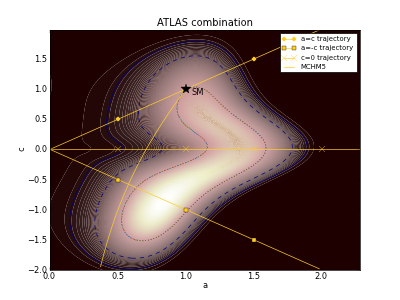}
\hfill
\end{minipage}
\caption{
{\it
ATLAS constraints on the couplings $(a, c)$ of a possible Higgs-like particle $h$ with mass $\sim 125$~GeV
arising from the (top left) ${\bar b} b$, (top right) $\tau^+ \tau^-$,
(centre left)  $Z Z^*$, (centre right) $W W^*$ and (bottom left) $\gamma \gamma$ final states. 
The bottom right panel displays the combination of these ATLAS constraints.}} 
\label{fig:ATLAS} 
\end{figure}

The six panels of Fig.~\ref{fig:ATLAS} display the corresponding ATLAS constraints
from the (top left) ${\bar b} b$, (top right) $\tau^+ \tau^-$,
(centre left)  $Z Z^*$, (centre right) $W W^*$ and (bottom left) $\gamma \gamma$ final states. Also shown in
the bottom right panel is the combination of these ATLAS constraints. We see again in the top panels that the
${\bar b} b$ constraint depends on both $a$ and $c$ whereas the $\tau^+ \tau^-$ measurements are 
mainly sensitive to $c$, as one would expect,
with a very weak dependence on $a$ induced by the subdominant vector-boson-fusion (VBF)
production mechanism. On the other hand, we see in the centre panels 
that the $Z Z^*$ and $WW^*$ channels are quite sensitive to both $a$ and $c$,
but are insensitive to the sign of $c$, whereas in the bottom left panel the constraint from the
$\gamma \gamma$ channel differs between positive and negative $c$. 
The combination of ATLAS constraints shown in the bottom right
panel of Fig.~\ref{fig:ATLAS} has a characteristic `boomerang' shape, reflecting the fact that
the data available from ATLAS provide no discrimination between the $gg$ fusion and VBF
production mechanisms.

\begin{figure}
\vskip 0.5in
\vspace*{-0.75in}
%\hspace*{-.70in}
\begin{minipage}{8in}
\includegraphics[height=6cm]{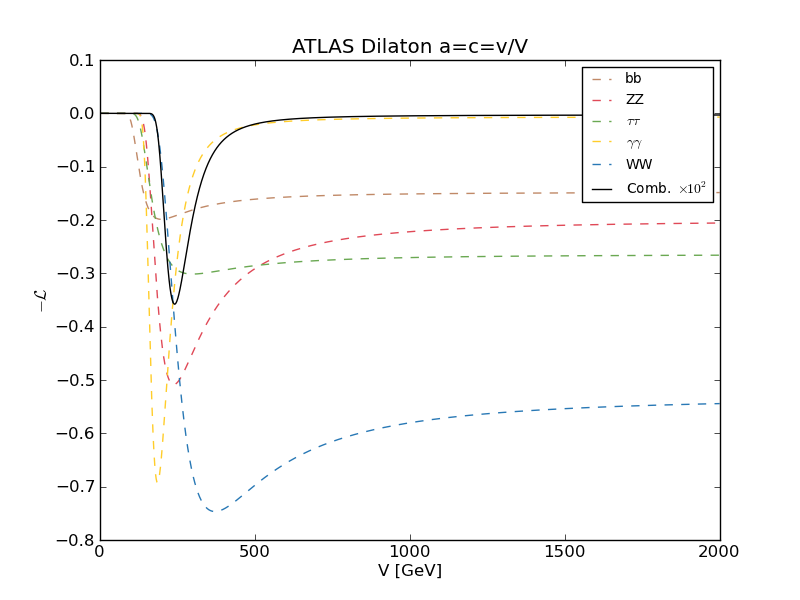}
\hspace*{-0.17in}
\includegraphics[height=6cm]{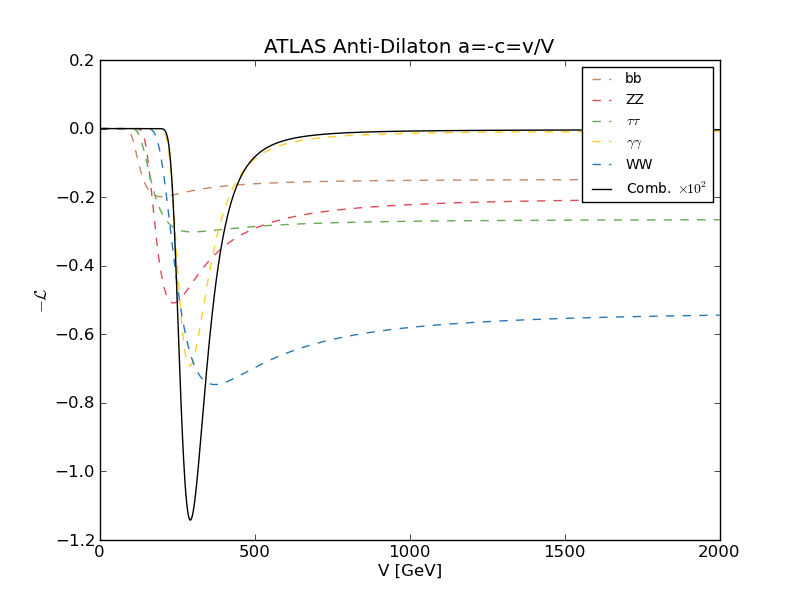}
\hfill
\end{minipage}
\begin{minipage}{8in}
\includegraphics[height=6cm]{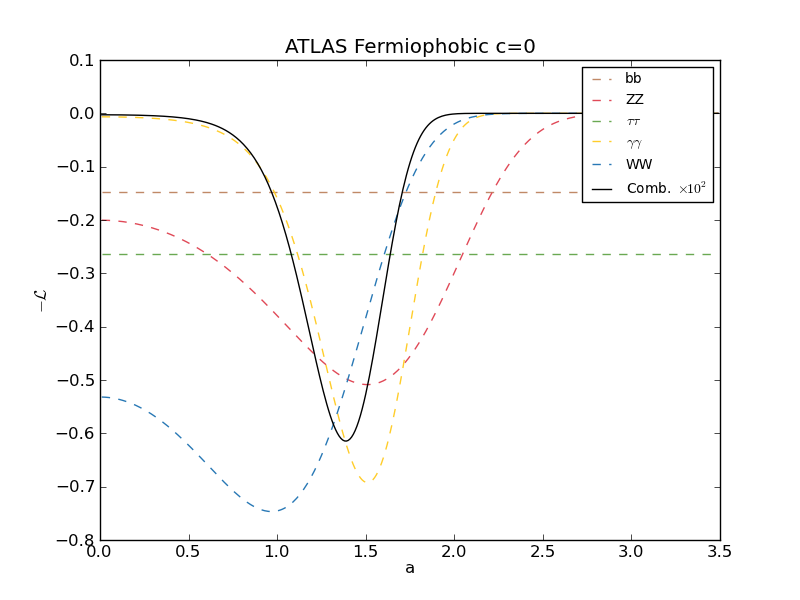}
\hspace*{-0.17in}
\includegraphics[height=6cm]{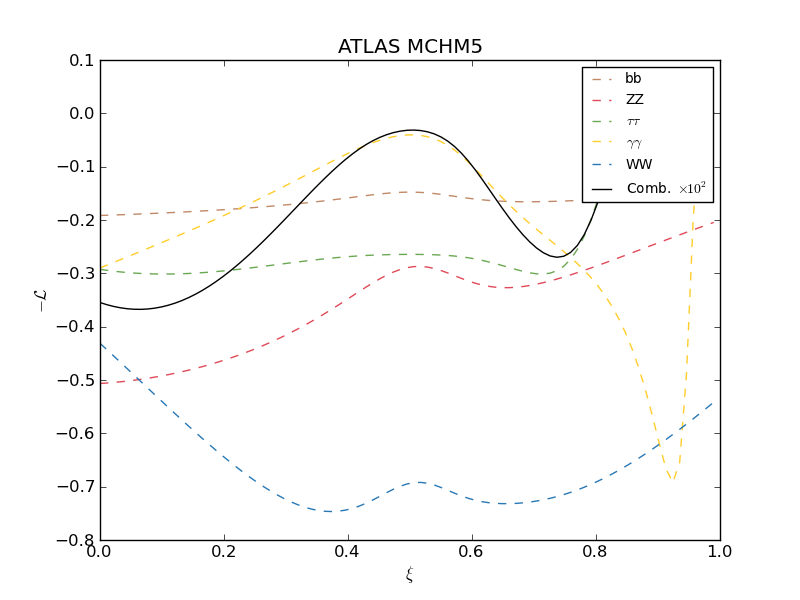}
\hfill
\end{minipage}
\caption{
{\it
The $p$-values of global fits to the ATLAS constraints along (upper left) the pseudo-dilaton line $a = c$,
(upper right) the `anti-dilaton' line $a = -c$, (lower left) the fermiophobic line $c = 0$, and
(lower right) the MCHM5 line $a c = 1 - 2\sqrt{1 - a^2}$.
}}
\label{fig:ATLASpvalues} 
\end{figure}

The $p$-values for the ATLAS data along the same specific model lines are shown in Fig.~\ref{fig:ATLASpvalues}.
We again see that the $p$-value along the pseudo-dilaton line is minimized for $a = c = v/V$
close to unity (the Standard Model case), and the $p$-value is somewhat improved for the 
anti-dilaton scenario. There is also a significant minimum of the $p$-value in the fermiophobic
scenario, which was disfavoured in the CMS data by measurements of the VBF-enhanced 
$W W^*$ and $\gamma \gamma$ sub-channels.

\begin{figure}
\vskip 0.5in
\vspace*{-0.75in}
%\hspace*{-.70in}
\begin{minipage}{8in}
\includegraphics[height=4cm]{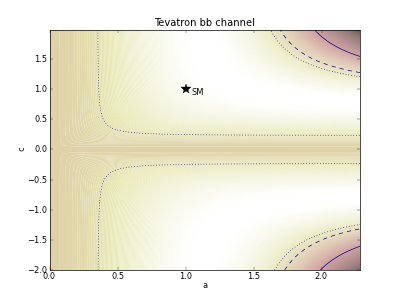}
%\hspace*{-0.17in}
\includegraphics[height=4cm]{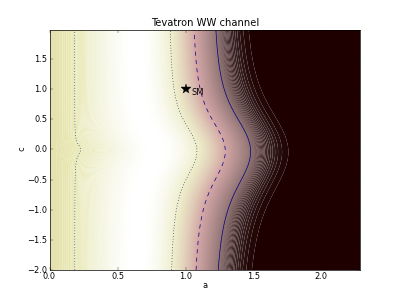}
%\hspace*{-0.17in}
\includegraphics[height=4cm]{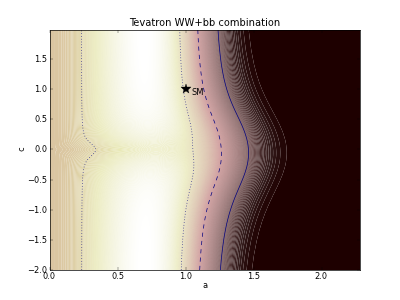}
\hfill
\end{minipage}
\caption{
{\it
The Tevatron constraints on the couplings $(a, c)$ of a possible Higgs-like particle $h$ with mass $\sim 125$~GeV
arising from the (left) ${\bar b} b$ and (centre) $WW^*$ final states, and (right) their combination.}} 
\label{fig:Tevatron} 
\end{figure}

Finally, in Fig.~\ref{fig:Tevatron} we display the constraints in the $(a, c)$ plane
provided by measurements by the Tevatron experiments, CDF and D0, 
in the (left) ${\bar b} b$ and (centre) $WW^*$
channels. We see that the ${\bar b} b$
channel disfavours fermiophobic scenarios, as seen in the combination
that is displayed in the right panel.

\begin{figure}
\vskip 0.5in
\vspace*{-0.75in}
%\hspace*{-.70in}
\begin{minipage}{8in}
\hspace*{-0.7in}
%\epsfig{file=Tevatron_WW.eps,height=1.6in}
%\hspace*{-0.17in}
\centerline{\includegraphics[height=8cm]{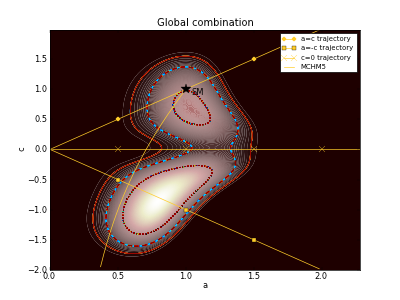}}
\hfill
\end{minipage}
\caption{
{\it
The constraints on the couplings $(a, c)$ of a possible `Higgs' $h$ particle with mass $\sim 125$~GeV
obtained from a global analysis of the available CMS, ATLAS, CDF and D0 data.}} 
\label{fig:Combined} 
\end{figure}

We discuss finally the combination of the above results.
Fig.~\ref{fig:Combined} shows the constraints in the $(a, c)$ plane obtained from a global
analysis of CMS, ATLAS and Tevatron data. We see two preferred regions of parameter space,
one for $c > 0$ and the other (which is somewhat favoured) for $c < 0$. The Standard Model
point lies just outside the 68\% CL contour for $c > 0$ (corresponding also to the MCHM5 model in the
limit $\xi \to 0$). The fermiophobic models are disfavoured well below the 68\% CL, primarily on the basis
of the CMS VBF-enhanced $WW^*$ and $\gamma \gamma$ sub-channels discussed above.

\begin{figure}
\vskip 0.5in
\vspace*{-0.75in}
%\hspace*{-.70in}
\begin{minipage}{8in}
\includegraphics[height=6cm]{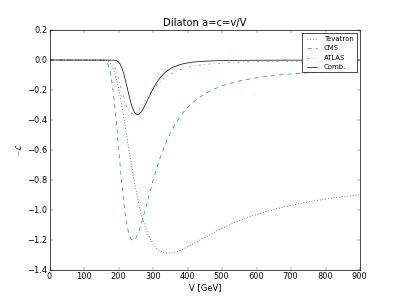}
\hspace*{-0.17in}
\includegraphics[height=6cm]{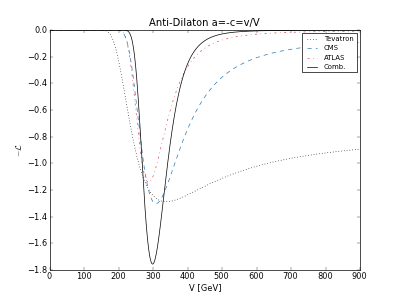}
\hfill
\end{minipage}
\begin{minipage}{8in}
\includegraphics[height=6cm]{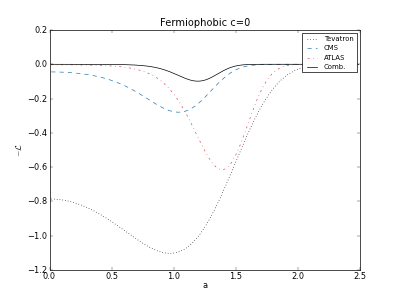}
\hspace*{-0.17in}
\includegraphics[height=6cm]{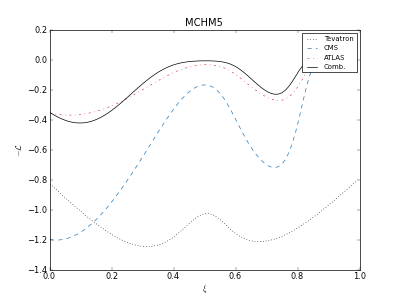}
\hfill
\end{minipage}
\caption{
{\it
The $p$-values of global fits to the CMS, ATLAS and Tevatron constraints along (upper left) the pseudo-dilaton line $a = c$,
(upper right) the `anti-dilaton' line $a = -c$, (lower left) the fermiophobic line $c = 0$, and
(lower right) the MCHM5 line $a c = 1 - 2\sqrt{1 - a^2}$.
}}
\label{fig:pvalues} 
\end{figure}

The same features are visible in Fig.~\ref{fig:pvalues}, where we see how the preference for the `anti-dilaton'
scenario arises. As previously, the upper
left panel is for the pseudo-dilaton scenario with $a = c$, the upper right panel is
for the anti-dilaton scenario with $a = -c$, the lower left panel is for fermiophobic models,
and the lower right panel is for the MCHM5
model. In the case of the pseudo-dilaton scenario (which includes the Standard Model and the
$\xi \to 0$ limit of the MCHM5 model when $V = 246$~GeV), we see that the values of $V$ favoured
by the CMS, ATLAS and Tevatron data, while overlapping, do not coincide, whereas they coincide perfectly
in the `anti-dilaton' case. The preference for $a/c < 0$ can be traced to the fact that both CMS and
ATLAS see $\gamma \gamma$ signals that are somewhat enhanced compared to the Standard Model, 
which can be explained by positive interference between the top and $W^\pm$ loops if $a$ and $c$ have
opposite signs, as seen in (\ref{loopfactors}).
In the fermiophobic scenario, we see that the values of $a$ preferred by the different
experiments also do not coincide well. The local minima of the $p$-value for the MCHM5 scenario
reflect those seen already for the dilaton and anti-dilaton scenarios~\footnote{We do not display the corresponding $p$-value scan for
a global fit to the gaugephobic scenario with $a = 0$, which is essentially featureless with no preferred
range of $c$.}.

\begin{figure}
\vskip 0.5in
\vspace*{-0.75in}
%\hspace*{-.70in}
\begin{minipage}{8in}
\includegraphics[height=6cm]{GLOBAL_LH_dilaton.png}
\hspace*{-0.17in}
\includegraphics[height=6cm]{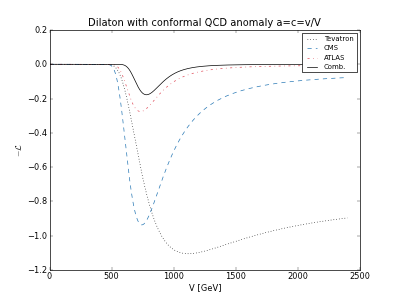}
\hfill
\end{minipage}
\begin{minipage}{8in}
\includegraphics[height=6cm]{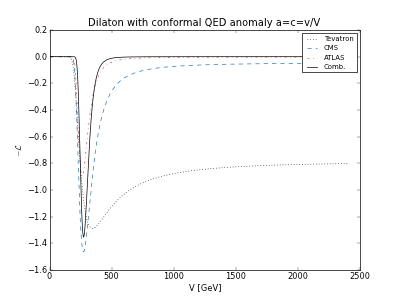}
\hspace*{-0.17in}
\includegraphics[height=6cm]{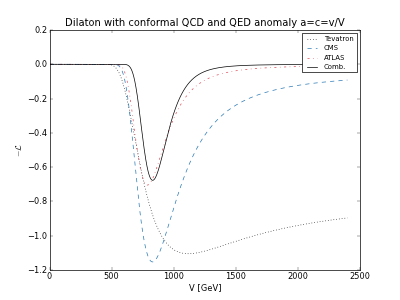}
\hfill
\end{minipage}
\caption{
{\it
The $p$-values of global fits to the CMS, ATLAS and Tevatron constraints along the pseudo-dilaton line $a = c$
(upper left) with only the Standard Model contributions to $b{s, em}$, (upper right) assuming that QCD is included
in the conformal sector, (lower left) assuming that QED is included in the conformal sector, and (lower right)
assuming that both QCD and QED are included in the conformal sector.
}}
\label{fig:pvalueswithanomalies} 
\end{figure}

Finally, we present in Fig.~\ref{fig:pvalueswithanomalies} the $p$-values for global fits in the
pseudo-dilaton model with $a = c = v/V$, varying assumptions on the nature of the
(near-)conformal sector. The upper left panel simply reproduces the upper left panel
of Fig.~\ref{fig:pvalues}, in which the possible contributions of the conformal sector to
the QCD loops for the $gg$ coupling and to the QED loops for the $\gamma \gamma$
coupling are ignored. We see that values of $V \sim v$ are favoured in this case. 
If QCD is included in the conformal sector as seen in the upper right panel of
Fig.~\ref{fig:pvalueswithanomalies}, enhancing $b_s$, larger values of $V \sim 800$~GeV are preferred, and the quality of
the best fit is somewhat reduced. On the other hand, if QED is included in the conformal 
sector as seen in the lower left panel of Fig.~\ref{fig:pvalueswithanomalies}, the preferred
range of $V \sim 250$~GeV as previously, and the quality of the best fit is significantly
improved. Finally, if both QCD and QED are included in the conformal sector, as seen
in the lower right panel of Fig.~\ref{fig:pvalueswithanomalies}, larger values of $V \sim 800$~GeV are again preferred
and the fit quality is improved relative to the case in which only QCD is conformal. These features
can easily be understood: the larger value of $V$ in the conformal QCD cases is because the data
are consistent with the Standard Model rate for $gg$ production of $h$, so the larger value of $b_s$
must be offset by a smaller value of $c = v/V$, and the improvements if QED is conformal arise from
the facts that both CMS and ATLAS see rates for $h \to \gamma \gamma$ that are somewhat enhanced
relative to the Standard Model.

\section{Summary and Prospects}

In this paper we have performed a global analysis of the data pertaining to a possible Higgs-like state $h$
with mass $\sim 125$~GeV that were made available by the CMS, 
ATLAS, CDF and D0 collaborations~\cite{LHCcombined,ZZsearch, bbbarsearch, CMSdiphotonsearch, ATLASdiphotonsearch, ATLASWWsearch, CMSWWsearch, CMSATLAStautausearch, CMStautaumumu, CMSWHsearch,CMSprevdiphotonsearch}
at the Moriond 2012 conference~\cite{Moriond}. 
We have considered a two-dimensional parameter space
characterized by rescaling factors $(a, c)$ for the possible $h$ couplings to massive vector bosons
and fermions, respectively~\cite{ac,Contino,mh125papers}. We have also considered the potential impacts of modifications to the
loop-induced $gg$ and $\gamma \gamma$ couplings that might be induced by additional massive
particles, specifically possible conformal QCD and/or QED sectors in the pseudo-dilaton scenario~\cite{pseudoDG}.

Good fits are found for models with $(a, c) \sim (1, 1)$ as in the Standard Model, pseudo-dilaton models
in which neither QCD nor QED is conformal and $V \sim v$, and the MCHM5 in the limit $\xi \to 0$. 
However, the fact that the CMS and ATLAS $\gamma \gamma$ signals are relatively large tends to favour
scenarios with $a/c < 0$. The same tendency has been found
in other phenomenological analyses, though the shapes of the preferred regions were somewhat different. This could
be expected, since phenomenological fits are necessarily approximate in the absence of more complete
information on the experimental likelihood functions. As we have discussed, fermiophobic scenarios are 
disfavoured by the CMS data on $WW^*$ and $\gamma \gamma$ dijet sub-channels, as also found in~\cite{Contino}.
In the pseudo-dilaton model, values of $V \sim 800$~GeV are preferred if QCD is (near-)conformal, and better fits are found 
if QED is (near-)conformal because the rate for $h \to \gamma \gamma$ is enhanced relative to
other decay modes.

We expect that the LHC will soon provide CMS and ATLAS with many more collisions, which
should make it possible to clarify the existence and nature of the possible Higgs-like state $h$ 
with mass $\sim 125$~GeV that has been reported. Within each experiment, this clarification will 
surely proceed via an analysis of the type discussed here, in which the signals from different channels
are combined statistically. Our analysis has demonstrated the value of the complementarity between the
different production and decay channels, and has highlighted the usefulness of information on
sub-channels corresponding to different production mechanisms. Above and beyond the analyses
made by individual collaborations, we hope that they will provide information on the likelihood
functions for different channels that will enable optimal global combinations to be made, going
beyond the crude approximations made here. This would hasten Judgement Day for the `Higgs' boson.

\section*{Acknowledgements}
The work of J.E. was supported partly by the London
Centre for Terauniverse Studies (LCTS), using funding from the European
Research Council via the Advanced Investigator Grant 267352. 
The work of T.Y. was supported by an STFC studentship.

\end{document}